# PPMXL and Gaia Morphological Analysis of Melotte 22 (Pleaides) and 25 (Hyades)

**W. H. Elsanhoury[1, 3] and M. I. Nouh[2, 3]**

[1] Physics Dept., Faculty of Science and Arts, Northern Border University, Rafha Branch, Saudi Arabia.
   Email: elsanhoury@nbu.edu.sa; welsanhoury@gmail.com
[2] Physics Dept., Faculty of Science, Northern Border University, Arar, Saudi Arabia.
   Email: nouh@nbu.edu.sa; abdo_nouh@hotmail.com
[3] Astronomy Dept., National Research Institute of Astronomy and Geophysics (NRIAG), 11421, Helwan, Cairo, Egypt

**Abstract**

In this article, we present a morphological study of the well-known extended sized open clusters in the sky, Melotte 22 (Pleaides) and Melotte 25 (Hyades) with J, H, and $K_s$ regions due to Gaia DR2 and PPMXL catalogues. Based on proper motions of these clusters, we extract our candidates and hence constructing fitting isochrones with Solar metallicity Z = 0.019, log (age) = 8.2 ± 0.05 for Melotte 22, and metallicity Z = 0.024, log (age) = 8.9 ± 0.10 for Melotte 25. Some photometric parameters are estimated, e.g. cluster heliocentric distances are 135 ± 3.6 and 47.51 ± 2.15 pc respectively, luminosity and mass functions with mass-luminosity relation MLR of the second order polynomial, and we have estimated the masses of Melotte 22 and Melotte 25 as 662.476 ± 25.73 $M_\odot$ and 513.819 ± 22.65 $M_\odot$. The relaxation times of these clusters are smaller than the estimated cluster ages, which indicate that these clusters are dynamically relaxed.

The second part is devoted to some kinematics of these clusters due to internal and/or motion of members, the apex position is 95°.60 ± 0°.10, -48°.21 ± 0°.14 for Melotte 22, and 96°.72 ± 1°.15, 5°.66 ± 0°.12 for Melotte 25. Also we have determined the components of space velocity and the cluster centers in pc as well as elements of the Solar motion. The results are in good agreement with those in the literature.

# 1. Introduction

Melotte 22 (Pleiades; NGC 1432; M45; Seven Sisters) and Melotte 25 (Collinder 50; Hyades) are the most famous star clusters can be seen by the naked eye in the constellation Taurus, these clusters are located in Northern Hemisphere. Over the past century, the stellar content of these two clusters has been studied extensively.

Melotte 22 (hereafter Me22) ($\alpha_{2000} = 03^h 45^m 59^s.2$, $\delta_{2000} = 24^h 22^m 09^s$) has a spatial location in the Galaxy as ($l = 166^o.199$, $b = -23^o.489$), its age is found over a wide range from 77 Myr (Mermilliod, 1981) to 141 Myr with MWSC II data (Kharchenko et al. 2013), while its metallicity [Fe/H] ∼ 0.03 as reported by Boesgaard & Friel (1990) and Taylor (2008). On the other hand, there are several estimates of the heliocentric distance of the Me22, e.g. Galli et al. (2017) gave the distance as $134^{+2.9}_{-2.8}$ pc, 136.2 ± 1.2 pc given by Melis et al. (2014) using an absolute trigonometric parallax distance measurement with the help of the Very Long Baseline Interferometry VLBI. Groenewegen et al. (2007) estimated the distance using double stars orbital modeling that led to the cluster distance 138.0 ± 1.7 pc. Me22 is located within the thin disk during its movement in space and reaches a maximum distance above the Galactic plane $Z_{max.} = 0.08$ pc (Wu et al. 2009).

On the other hand, Melotte 25 (hereafter Me25) is one of the most famous open clusters of the Milky Way Galaxy with location of ($\alpha_{2000} = 04^h 26^m 52^s.0$, $\delta_{2000} = 15^h 52^m 01^s$) and ($l = 180^o.058$, $b = -22^o.349$). In addition, Me25 is the moderately rich cluster, with some 300 – 1000 possible members and an age of around 600 – 800 Myr (Perryman et al. 1998).

The importance of Me25 study regarding understands the chemical evolution of the Galaxy and demonstration of Galactic structure, as well as the determination of the population I distance scale. It has played a fundamental role in astronomy as a first step on the cosmic distance ladder and as a test case for theoretical models of stellar interiors.

The distance to the Me25 has for many years been based solely on the convergent point method until Hodge and Hodge (1966) suggested that the cluster was located some 20% farther from the Sun than indicated by the proper motion. van Altena (1974) reviewed the various methods used for determining the Me25 distance modulus, e.g. proper motions, trigonometric parallaxes of Me25 members, dynamical parallaxes, the Ca II K-line absolute magnitudes, photometric parallaxes and stellar-interior calculations. He concluded that all "secondary" distance indicators yield distance moduli greater than those determined from proper motion.

In our study, we use the fundamental parameters derived by Kharchenko et al. (2013 and 2016) and Dias et al. (2002) to determine the basic astronomical and photometrical properties of open clusters Me22 and Me25. We extract a complete worksheet data with aid of the second Gaia data DR2[1] release and PPMXL[2] catalogues. Gaia DR2 contains precise astrometry at the sub-milliarcsecond level and homogeneous photometry at the mag level, which could be used to characterize a large number of clusters over the entire sky. PPMXL catalogue (Roeser et al., 2010) contains the positions and proper motions of USNO-B1.0 and the Near Infrared photometry of the 2MASS. PPMXL contains a total number of about 900 million objects, some 410 million with 2MASS photometry, and in the largest collection of International Celestial Reference System ICRS of proper motion at present.

This paper is organized as follows. In section 2, we focused on the data analysis ends with the estimation of the photometric parameters. Section 3, deals with some of the kinematical properties of Me22 and Me25. Finally, the conclusion of the article is presented in Sect. 3.

---

[1] http://cdsarc.u-strasbg.fr (130.79.128.5) or via http://cdsweb.u-strasbg.fr/cgi-bin/qcat?J/A+A/618/A93
[2] http://vizier.cfa.harvard.edu/viz-bin/VizieR?-source=I/317

## 2. Data Analysis

In Table 1, we listed the fundamental parameters of Me22 and Me25 taken from Kharchenko et al. (2013 and 2016) and Dias et al. (2002).

Table 1: The fundamental parameters of two open clusters Me22 and Me25.

| Parameter | Me22 | Me25 | References |
|---|---|---|---|
| $\alpha$ | $03^h\ 45^m\ 59^s.2$ | $04^h\ 26^m\ 52^s.0$ | Kharchenko et al. (2016) |
| | $03^h\ 47^m\ 0^s.0$ | $04^h\ 26^m\ 54^s.0$ | Dias et al. (2002) |
| $\delta$ | $24^d\ 22^m\ 09^s$ | $15^d\ 52^m\ 1^s.0$ | Kharchenko et al. (2016) |
| | $24^d\ 07^m\ 00^s$ | $15^d\ 52^m\ 0^s.0$ | Dias et al. (2002) |
| $l$ | $166°.199$ | $180.°758$ | Kharchenko et al. (2016) |
| | $166°.571$ | $180°.064$ | Dias et al. (2002) |
| $b$ | $-23°.489$ | $-22°.349$ | Kharchenko et al. (2016) |
| | $-23°.521$ | $-22°.064$ | Dias et al. (2002) |
| Distance $_{pc}$ | 130 | 50 | Kharchenko et al. (2016) |
| | 133 | 45 | Dias et al. (2002) |
| E(B-V) $_{mag}$ | 0.021 | 0.030 | Kharchenko et al. (2016) |
| | 0.021 | 0.010 | Dias et al. (2002) |
| log (age) $_{yr}$ | 8.150 | 8.896 | Kharchenko et al. (2016) |
| | 8.131 | 8.896 | Dias et al. (2002) |
| (m-M) $_{mag}$ | 5.58 | 3.48 | Kharchenko et al. (2016) |
| Diameter $_{arcmin}$ | 120 | 330 | Dias et al. (2002) |
| Z (metallicity) | 0.019 | - | Tadross et al. (2010) |
| | - | 0.024 | Perryman et al. (1998) |

### 2.1 Cluster Center Determination

Using Gaia DR2 database service (http://cdsweb.u-strasbg.fr/cgi-bin/qcat?J/A+A/618/A93), we downloaded the data within a radius of 60 and 165 arcmin of Me22 (26,527 points) and Me25 (191,234 points). To determine the clusters centers (the location of the maximum stellar density of the cluster's area), we use TOPCAT software. We follow the procedure presented by many authors, e.g. Maciejewski and Niedzielski (2007), Maciejewski et al. (2009), and Haroon et al. (2014 and 2017), into which two perpendicular strips were cut along the right ascension and declination at approximate center of the cluster, and then the Histograms of the star counts was built along each strip with bin size of 1 arcmin for the clusters with a diameter larger than 10 arcmin (Maciejewski and Niedzielski 2007), which fitted by Gaussian distribution function *f(x)* with mean µ and standard deviation σ, i.e.

$$f(x) = \frac{1}{\sigma\sqrt{2\pi}} e^{-(x-\mu)^2/2\sigma^2},$$

where the location of a maximum number of stars (peak) indicates the new cluster center. Figure 1 shows the new clusters centers and the fitting parameters are listed in Table 2.

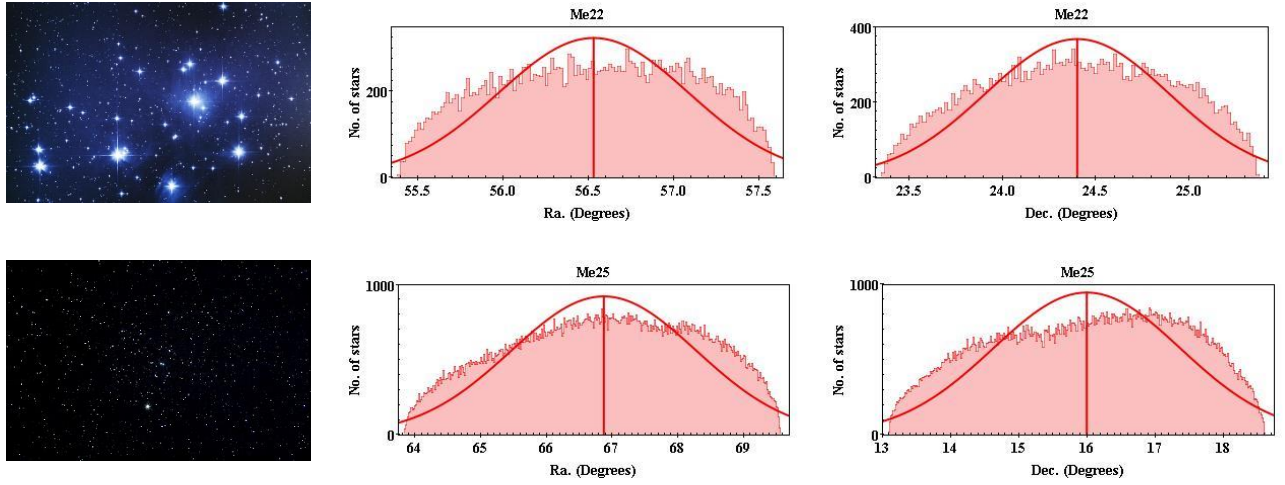

**Fig. 1:** *Upper panel:* presents an image of the Me22 open cluster with new center determination by TOPCAT. *Lower panel* presents an image of the Me25 open cluster with new center determination also by TOPCAT.

By comparing our results with the authors listed in Table 1, we notice that:

- For Me22, our new estimation of right ascension is greater than the value of Kharchenko et al. (2016) by about $8^s.3$ and less than the value of Dias et al. (2002) by about $1^m 0^s.8$. On the other hand, our estimated declination is greater than both values of Kharchenko et al. (2016) by about $1^m 55^s.79$ and by about $17^m 4^s.79$ of Dias et al. (2002).

- For Me25, our new estimation of right ascension is greater than the value of Kharchenko et al. (2016) by about $38^s.24$ and by about $36^s.24$ for Dias et al. (2002). On the other hand, our estimated declination is greater than both values of Kharchenko et al. (2016) by about $7^m 55^s.4$ and by about $7^m 56^s.4$ of Dias et al. (2002).

| Parameter | Me22 | Me25 |
|---|---|---|
| Max. peak (Ra.) degrees | 56.531 ± 0.557 | 66.876 ± 1.411 |
| Max. peak (Dec.) degrees | 24.401 ± 0.491 | 15.999 ± 1.373 |
| α | 03$^h$ 46$^m$ 7$^s$.5 | 04$^h$ 27$^m$ 30$^s$.24 |
| δ | 24$^d$ 24$^m$ 4$^s$.79 | 15$^d$ 59$^m$ 56$^s$.4 |
| l$^o$ | 166°.201 | 180°.051 |
| b$^o$ | -23°.444 | -22°.147 |

Table 2: Our center's estimation of Me22 and Me25.

## 2.2 Radial Density Profile RDP

Due to the internal and/or external dynamical process taking place in and out of the cluster, it is important to study the distribution of the mean surface density ρ(r) in concentric rings as a function of radius from the cluster center outward, which is referred to radial density profile RDP. Using our calculated values of the new center (α, δ), the new worksheet data with Gaia DR2 catalogue contains right ascension, declination, and the angular distance from the cluster center.

The density ρ(r) of each ring is estimated by dividing the number of stars in the ring by its area (i.e. $N_i / A_i$), and use the empirical relation of King (1966), which parameterize ρ(r) as

$$\rho(r) = f_{bg} + \frac{f_0}{1+(r/r_{core})^2}, \qquad (1)$$

where $r_{core}$, $f_o$, and $f_{bg}$ are the core radius, the central surface density, and the background surface density, respectively. In addition, we shall define the limiting radius $r_{lim}$ as the radius which covers the entire cluster area and reaches enough stability with the background field density (Tadross and Bendary, 2014). Mathematically, $r_{lim}$ is given by

$$r_{lim} = r_{core} \sqrt{\frac{f_\circ}{3\sigma_{bg}} - 1}. \qquad (2)$$

where $\sigma_{bg}$ is the uncertainty of the background surface density $f_{bg}$.

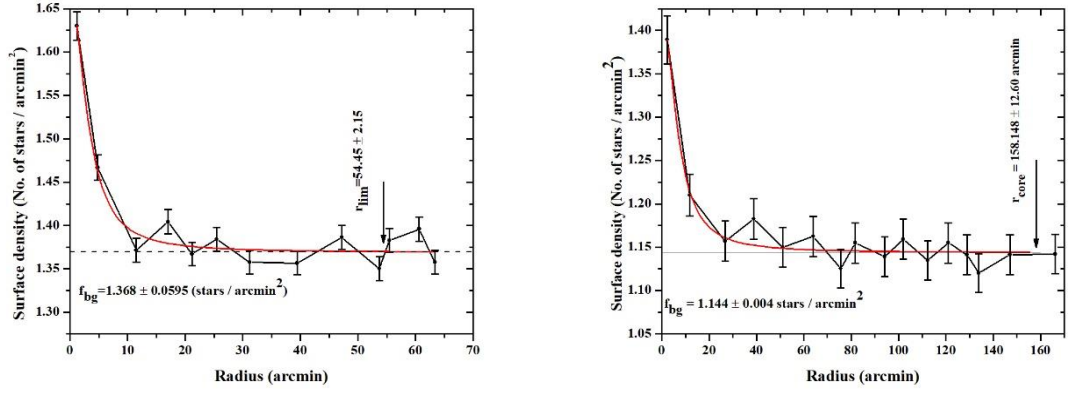

**Fig. 2:** RDP of the Me22 *(left panel)* and Me25 *(right panel)* open clusters with error bars, the fitted solid lines denote the density distribution and the dashed lines represent the background field density $f_{bg}$.

Fig. 2, represents our calculations for the surface density distribution $\rho(r)$. The numerical values of $r_{core}$, $f_{bg}$, and $f_o$ are listed in Table 3. Where, our calculated $r_{core}$ for Me22 is less than that given by Kharchenko et al. (2013) by about 0.81 pc, while for Me25 our estimated value is smaller than that of Röser et al. (2011), Perryman et al. (1998), and Gunn et al. (1988) by about 0.80 pc, 0.40 pc, and 0.85 pc, respectively.

Our calculations of the concentration parameter ($C = r_{lim}/r_{core}$) for both clusters are given also in Table 3. Nilakshi et al. (2002) concluded that the angular size of the coronal region is about $6r_{core}$, while Maciejewski and Niedzielski (2007) reported that, $r_{lim}$ ranged between $2r_{core}$ and $7r_{core}$. Our estimation of concentration parameter C is about $2.07 \pm 0.15$ and $4.53 \pm 0.03$ for Me22 and Me25 open clusters, respectively. Therefore, our calculation is in a good agreement with Maciejewski and Niedzielski (2007).

**Table 3:** Our estimated Me22 and Me25 RDP parameters.

| Parameter | Me22 | Me25 | References |
|---|---|---|---|
| $f_{bg}$ (stars / arcmin$^2$) | 1.325 ± 0.01 | 1.144 ± 0.004 | Present work |
| $f_o$ (stars / arcmin$^2$) | 0.159 ± 0.003 | 0.273 ± 0.020 | Present work |
| $r_{core}$ (pc) | 1.35 ± 0.01 | 2.30 ± 0.03 | Present work |
| | 1.3 – 2.1 | 3.1 | Fujii and Hori (2018) |
| | 2.16 | - | Kharchenko et al. (2013) |
| | - | 3.10 | Röser et al. (2011) |
| | - | 2.7 | Perryman et al. (1998) |
| | - | 3.15 | Gunn et al. (1988) |
| $r_{lim}$ (pc) | 2.80 ± 0.1 | 10.43 ± 1.15 | Present work |
| C | 2.07 ± 0.15 | 4.53 ± 0.03 | Present work |

## 2.3 Color-Magnitude Diagram CMD

Now both data of Gaia DR2 and PPMXL were cross (or X) matched using TOPCAT, getting 20,324 and 112,942 sources, which contains all the data we need, i.e. three colors J, H, and $K_s$ magnitudes with their proper motion $\mu_\alpha \cos\delta$ and $\mu_\delta$ (mas/yr). Stars with proper motion lie between ±1 sigma could be considered only as a membership probability, so we have 903 candidates for Me22 and 1712 candidates for Me25.

One of the main target of our study is to estimate the photometric parameters (e.g. reddening, distance modulus, … etc.) of Me22 and Me25 open clusters, which can be achieved by constructing the CMD and fitting it with the theoretical Padova isochrones[1], Marigo et al. (2008) and Girardi et al. (2010). Here, and based on this isochrone models, we fitted our CMD for (J, J-H & $K_s$, J- $K_s$) with Solar metallicity Z = 0.019, log (age) = 8.15, 8.20, and 8.25 yr for Me22 and Z = 0.024, log (age) = 8.8, 8.9, and 9.0 yr for Me25, the results are shown in Fig. 3.

The reddening of the cluster has been determined using the relations of Schlegel et al. (1998) Schlafly and Finkbeiner (2011). We have the coefficient ratios $A_J/A_V = 0.276$ and $A_H/A_V = 0.176$, which are derived using absorption ratios by Schlegel et al. (1998), while the ratio $A_{Ks}/A_V = 0.118$ was derived by Dutra et al. (2002). For our calculation we used the following values for the color excess by Fiorucci and Munari (2003) as: $E_{J-H}/E_{B-V} = 0.309 \pm 0.130$, $E_{J-K}/E_{B-V} = 0.485 \pm 0.150$, where $R_V = A_V/E_{B-V} = 3.1$. We have used these values of Me22 and Me25 to correct the effects of reddening in the CMDs with an extinction coefficient $A_V$ equal to 0.018 mag. and 0.005 mag. for Me22 and Me25, respectively. Our calculations indicate that the heliocentric distances r are $135 \pm 3.60$ pc and $47.51 \pm 2.15$ pc for Me22 and Me25, respectively.

---

[1] http://stev.oapd.inaf.it/cgi-bin/cmd

Comparison of our results for r, $E(B - V)$, and log (age), with that obtained by Kharchenko et al. (2016) and Dias et al. (2002) is listed in Table 1. We notice that for Me22, our distance is much greater by about 5 and 2 pcs than those obtained by Kharchenko et al. (2016) and Dias et al. (2002) respectively, while our $E(B - V)$ is greater by 0.015 mags. Also, our log (age) is greater by 0.05 and 0.069 yr than that due to Kharchenko et al. (2016) and Dias et al. (2002), respectively. For Me25, our estimated distance is smaller by about 2.49 pc than that obtained by Kharchenck et al. (2016) and greater than that obtained by Dias et al. (2002) by about 2.51 pc, while our $E(B - V)$ is smaller than Kharchenko et al. (2016) and Dias et al. (2002) values by about 0.02. Also for those same authors, our log (age) is smaller than by about 0.004 yr.

The estimated heliocentric distances led us to determine the cluster's distance to the Galactic center $R_{gc}$, the projected distance to the Galactic plane $X_\odot$ and $Y_\odot$, and the distance from the Galactic plane $Z_\odot$ (Tadross 2012), the results are presented in Table 4.

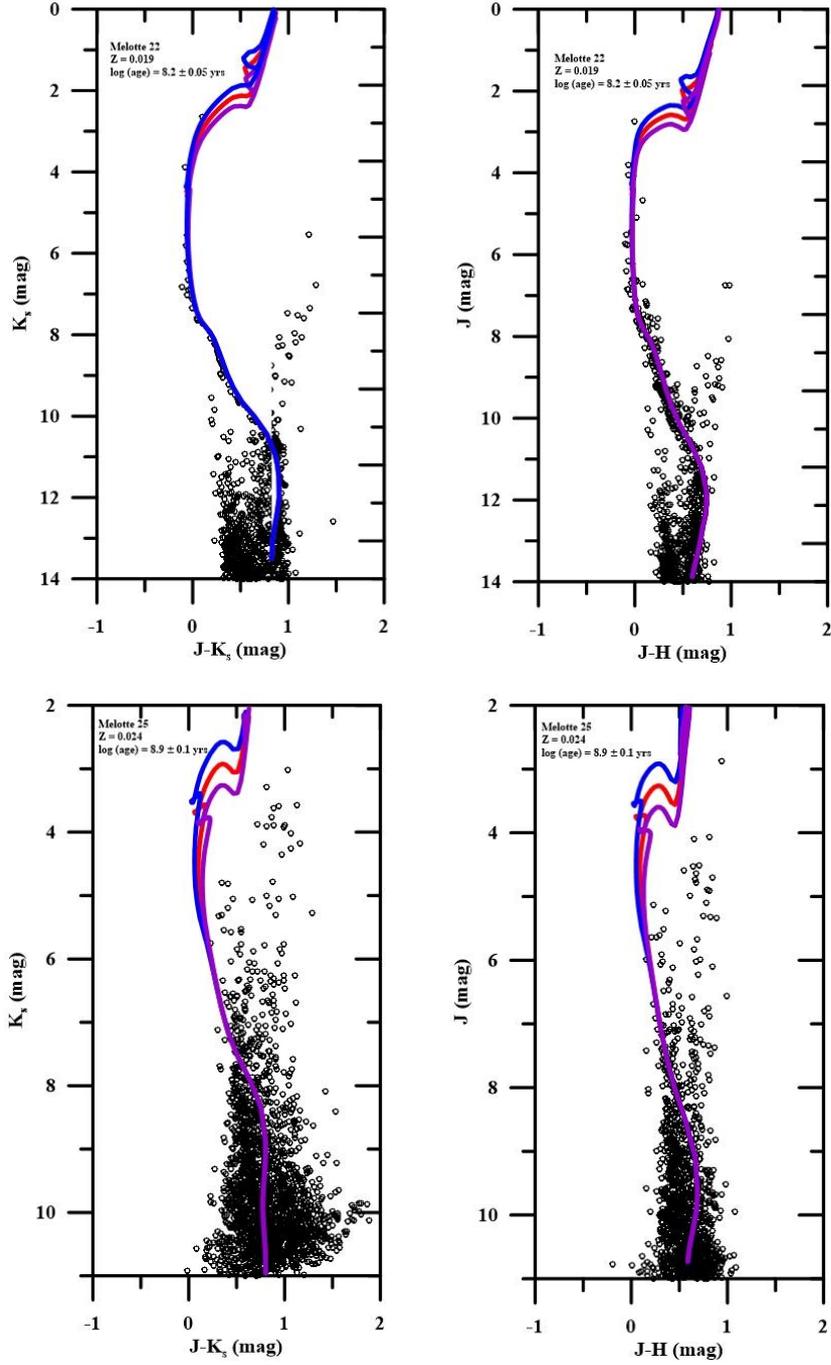

**Fig. 3:** Padova CMD over [J, (J-H) and ($K_s$, (J-$K_s$)] isochrones for Me22 *(upper panel)* and Me25 *(lower panel)*.

## 2.4 Luminosity and Mass Functions

The property used to study large groups or classes of objects like stars in clusters and/or galaxies is called the luminosity function LF, which describes the density of stars in different absolute magnitudes (Haroon et al. 2017). Fig. 4 illustrates the LF of Me22 and Me25 clusters.

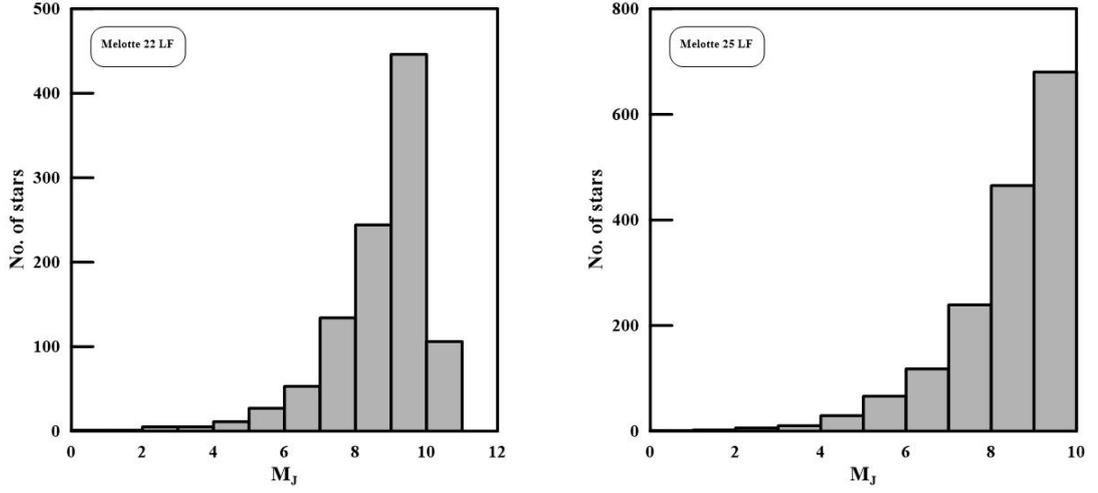

**Fig. 4:** The LF of Me22 *(left panel)* and Me25 *(right panel)*.

Here we analyze the observed stars counts as a function of magnitude to obtain the LF and mass function MF, as well as their spatial dependence by their mass-luminosity relation MLR. Scalo (1986) defined the initial mass function as an empirical relation that describes the mass distribution (i.e. Histograms of stellar masses) of a population of stars as a function of their theoretical initial mass, i.e.

$$\frac{dN}{dM} \propto M^{-\Gamma} \quad , \tag{3}$$

where dN/dM is the number of stars on the mass interval (M : M + dM), and $\Gamma$ is a dimensionless exponent. From Salpeter (1955), the IMF for massive stars (> 1$M_\odot$) has been studied and well established with $\Gamma = 2.35$.

In this context, MF could be obtained from LF using Padova theoretical evolutionary tracks and their isochrones. The relation is a polynomial function of the second order as

$$\frac{M}{M_\odot}(\text{Me22}) = 3.364063993 - 0.9090064094\, M_K + 0.06665147753\, M_K^2,$$
$$\frac{M}{M_\odot}(\text{Me25}) = 2.41391221 - 0.4946243929\, M_K + 0.02741943709\, M_K^2. \tag{4}$$

Fig. 5 shows the MF's of these clusters where the slope is determined as 2.38 ± 0.65 and 2.16 ± 0.68, for Me22 and Me25 respectively, which are in a good agreement with the Salpeter's values.

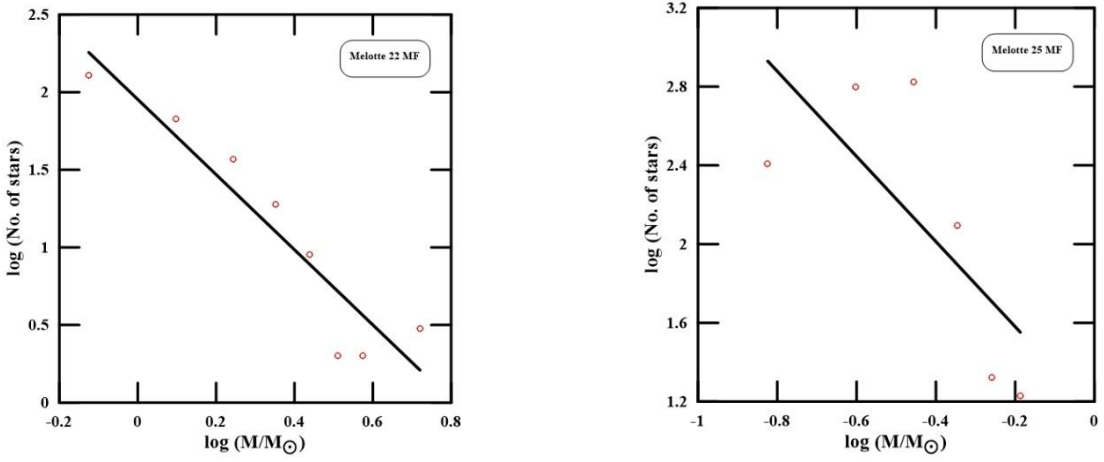

**Fig. 5:** The MF of Me22 *(left panel)* and Me25 *(right panel)*. The slopes are 2.38 ± 0.65 and 2.16 ± 0.68, respectively.

By using our total estimated mass $M_C$, we can determine the tidal radius $r_t = 1.46\sqrt[3]{M_c}$ (Jeffries et al. 2001) of these objects, the results are listed in Table 4.

**2.5 Dynamical State of Me22 and Me25**

The time needed for the cluster to reach stability due to contraction and destruction forces (Maxwellian equilibrium) is called the relaxation time $T_{relax}$. In this time the low mass star in a cluster possesses the largest random velocity, occupying a larger volume than the high mass does (Mathieu and Latham, 1986). $T_{relax}$ depending mainly on the number N of members and the cluster diameter. Mathematically $T_{relax}$ has the form (Maciejewski and Niedzielski, 2007).

$$T_{relax} = \frac{N}{8\ln N} T_{cross}. \qquad (5)$$

where $T_{cross}$ is the crossing time, i.e. the time needed where the central part of the cluster to be relaxed. In this context, the crossing time of Me22 is about 8 Myr (Pinfield et al., 1998) and for Me25 is 30 Myr and 20 Myr (Pels et al. 1975) and de Bruijne et al. (2001) respectively. Finally, we evaluate the dynamical evolution parameter $\tau = T_{age}/T_{relax}$, which describes the dynamical state of the cluster. Table 4 shows the numerical values of these parameters.

## 3. Kinematical Analysis

The stars included in open clusters share similar parameters, e.g. distance, age, chemical composition, and hence kinematics. Kinematics of the two clusters under considerations will be estimated according to the algorithms discussed in the next subsections, Elsanhoury et al. (2015 and 2018).

### 3.1 Components of Space Velocity

For a group of N cluster member stars with coordinates ($\alpha$, $\delta$) located at a distance $r_i$ (pc), proper motions in RA and DEC, i.e. $\mu_\alpha \cos\delta$ and $\mu_\delta$ (mas/yr) and radial velocity $V_r$ (km/s). In this manner, the velocity components $V_x$, $V_y$, and $V_z$ along x, y, and z-axes in the coordinate system centered at the Sun presented by the well-known formulae by Smart (1968),

$$V_x = -4.74 r_i \mu_\alpha \cos\delta \sin\alpha - 4.74 r_i \mu_\delta \sin\delta \cos\alpha + V_r \cos\delta \cos\alpha, \quad (6)$$

$$V_y = +4.74 r_i \mu_\alpha \cos\delta \cos\alpha - 4.74 r_i \mu_\delta \sin\delta \sin\alpha + V_r \cos\delta \sin\alpha, \quad (7)$$

$$V_z = +4.74 r_i \mu_\delta \cos\delta + V_r \sin\delta. \quad (8)$$

We used our estimated heliocentric distances $r_i$ of the star clusters under considerations, i.e. 135 ± 3.6 and 47.51 ± 2.15 pc for Me22 and Me25, respectively and the radial velocities $V_r$ of the Me22 as 3.503 ± 0.391 km s$^{-1}$ from RAVE catalogue (Conrad et al. 2014), and 39.01 ± 0.02 km s$^{-1}$ (David et al. 2018) for Me25.

In order to compute the components of space velocity U, V, and W along galactic space coordinates, we have used the transformations due to Murry (1989)

$$U = -0.054875539\ V_x - 0.873437105\ V_y - 0.483834992\ V_z, \quad (9)$$

$$V = \phantom{-}0.494109454\ V_x - 0.444829594\ V_y + 0.746982249\ V_z, \quad (10)$$

$$W = -0.867666136\, V_x - 0.198076390\, V_y + 0.455983795\, V_z. \qquad (11)$$

while the mean velocities are given by

$$\bar{U} = \frac{1}{N}\sum_{i=1}^{N} U_i, \quad \bar{V} = \frac{1}{N}\sum_{i=1}^{N} V_i, \quad \bar{W} = \frac{1}{N}\sum_{i=1}^{N} W_i. \qquad (12)$$

### 3.2 Vertex, Center and Solar Motion

In order to calculate the apex (i.e. vertex) of the cluster, we could use the AD-chart method discussed by Chupina et al. (2001 and 2006) and Elsanhoury et al. (2016 and 2018), where the apex of individual stars is plotted on the so-called AD-diagram. In this method, equatorial coordinates of the convergent point are calculated from

$$A_{conv} = \tan^{-1}\left(\bar{V}_y / \bar{V}_x\right), \qquad (13)$$

$$D_{conv} = \tan^{-1}\left(\bar{V}_z / \sqrt{\bar{V}_x^2 + \bar{V}_y^2}\right). \qquad (14)$$

In Fig. 6 illustrate the AD-diagram for Me22 and Me25 clusters.

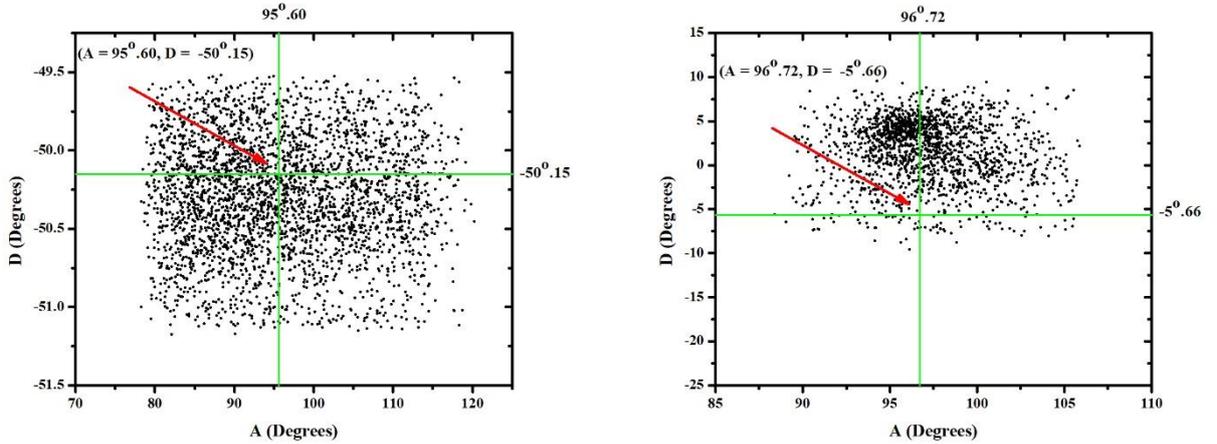

**Fig. 6:** The AD-charts for 903 Me22 *(left panel)* and 1712 Me25 *(right panel)*.

The cluster center $x_c$, $y_c$, and $z_c$ can be estimated by finding the equatorial coordinates of the center of mass for the number N

$$x_c = \left[\sum_{i=1}^{N} r_i \cos\alpha_i \cos\delta_i \right] \bigg/ N, \qquad (15)$$

$$y_c = \left[\sum_{i=1}^{N} r_i \sin\alpha_i \cos\delta_i \right] \bigg/ N, \qquad (16)$$

$$z_c = \left[\sum_{i=1}^{N} r_i \sin\delta_i \right] \bigg/ N. \qquad (17)$$

The Solar motion is given by

$$S_\odot = \left(\overline{U}^2 + \overline{V}^2 + \overline{W}^2\right)^{1/2}. \ km\,s^{-1}. \qquad (18)$$

Also, the galactic longitude $l_A$ and galactic latitude $b_A$ of the Solar apex are given by

$$l_A = tan^{-1}\left(-\overline{V}/\overline{U}\right), \qquad (19)$$

$$b_A = sin^{-1}\left(-\overline{W}/S_\odot\right), \qquad (20)$$

Table 4 listed the vertex, centers of the clusters and solar motion elements.

**Table 4:** The morphological analysis of Me22 and Me25 open clusters.

| Parameter | Me22 | Me25 | Reference |
|---|---|---|---|
| log (age) (yr) | $8.2 \pm 0.05$ | $8.9 \pm 0.10$ | Present work |
| | 8.15 | 8.962 | Kharchenko et al. (2016) |
| | 8.08 | - | Tetzlaff et al. (2010) |
| No. of members | 903 | 1712 | Present work |
| Metal abundance (Z) | 0.019 | 0.024 | Present work |
| | 0.019 | - | Tadross et al. (2010) |
| | - | 0.024 | Perryman et al. (1998) |
| E(B-V) | 0.036 | 0.010 | Present work |
| | 0.021 | 0.030 | Kharchenko et al. (2016) |
| | 0.021 | 0.010 | Dias et al. (2002) |
| E(J-$K_s$) | 0.017 | 0.002 | Present work |
| E(J-H) | 0.011 | 0.005 | Present work |
| (m-M) | $5.67 \pm 0.22$ | $3.89 \pm 0.13$ | Present work |
| r (pc) | $135 \pm 3.60$ | $47.51 \pm 2.15$ | Present work |
| | 130 | 50 | Kharchenko et al. (2016) |
| | $134^{+2.9}_{-2.8}$ | - | Galli et al. (2017) |
| Luminosity (mag.) | 1.633 | 2.291 | Present work |
| $X_\odot$ (kpc) | $-0.120 \pm 0.01$ | $-0.044 \pm 0.002$ | Present work |
| | $-8.12 \pm 0.024$ | - | Wu et al. (2009) |
| | $-0.116$ | $-0.046$ | Kharchenko et al. (2016) |
| $Y_\odot$ (kpc) | $0.030 \pm 0.02$ | $0.0004 \pm 0.001$ | Present work |
| | 0.030 | $-0.00005$ | Kharchenko et al. (2016) |
| | $0.028 \pm 0.006$ | - | Wu et al. (2009) |
| $Z_\odot$ (kpc) | $-0.054 \pm 0.02$ | $-0.018 \pm 0.001$ | Present work |
| | $-0.052$ | $-0.0014$ | Kharchenko et al. (2016) |
| | $-0.053 \pm 0.011$ | - | Wu et al. (2009) |
| $R_{gc}$ (kpc) | $8.5 \pm 0.01$ | 8.5 | Present work |
| | $8.1 \pm 0.0$ | - | Wu et al. (2009) |
| | - | 8.5 | Röser et al. (2011) |
| Total mass $M_c$ ($M_\odot$) | $662.476 \pm 25.73$ | $513.819 \pm 22.65$ | Present work |
| | 800 | - | Joseph et al. (2001) |
| | - | 435 | Röser et al. (2011) |
| Average mass $M_\odot$ | 0.734 | 0.300 | Present work |
| $r_t$ (pc) | $12.730 \pm 3.6$ | $11.694 \pm 3.42$ | Present work |
| | 16 – 19 | 10.5 | Fujii and Hori (2018) |
| | 16.53 | - | Kharchenko et al. (2013) |
| | - | 3.10 | Röser et al. (2011) |
| $\Gamma$ | $2.38 \pm 0.65$ | $2.16 \pm 0.68$ | Present work |
| $T_{relax}$ (Myr) | $132.70 \pm 11.52$ | $574.85 \pm 23.98$ (for $T_{cross} = 20$ Myr) | Present work |
| | - | $862.28 \pm 29.36$ (for $T_{cross} = 30$ Myr) | Present work |
| | 96 | - | Pinfield et al. (1998) |
| | 150 | 390 | Simon et al. (2001) |
| $\tau$ | 1.20 | 1.38 (for $T_{cross} = 20$ Myr) | Present work |
| | - | 0.92 (for $T_{cross} = 30$ Myr) | Present work |
| | 1.04 | - | Pinfield et al. (1998) |
| | 0.77 | 1.60 | Simon et al. (2001) |
| $(V_x, V_y, V_z)_{avg.}$ (km s$^{-1}$) | $-2.08, 21.23, -25.56$ | $-5.43, 46.08, 4.52$ | Present work |
| | $-1.07, 20.26, -23.26$ | - | Elsanhoury et al. (2018) |
| | - | $-6.23, 45.19, 5.31$ | Perryman et al. (1998) |
| $(U, V, W)_{avg.}$ (km s$^{-1}$) | $-6.06 \pm 0.41, -27.27 \pm 4.93, -13.75 \pm 0.27$ | $-42.14, -19.80, -2.35$ | Present work |
| | $-6.38 \pm 0.32, -26.91 \pm 2.04, -13.69 \pm 0.16$ | - | Elsanhoury et al. (2018) |
| | $-6.4 \pm 0.5, -24.4 \pm 0.7, -13.0 \pm 0.4$ | - | Galli et al. (2017) |
| | $-6.4 \pm 0.3, -26.8 \pm 0.1, -13.6 \pm 0.2$ | - | Tetzlaff et al. (2010) |
| $(A_{conv.}, D_{conv.})$ | $95°.60 \pm 0.10, -48°.21 \pm 0.14$ | $96°.72 \pm 1.15, 5°.66 \pm 0.12$ | Present work |
| | $95°.73 \pm 3.56, -50°.44 \pm 8.84$ | - | Elsanhoury et al. (2018) |
| | $92°.9 \pm 1.2, -49°.4 \pm 1.2$ | - | Galli et al. (2017) |
| | - | $97°.23 \pm 1°.41, 6°.96 \pm 0°.74$ | Vereshchagin et al. (2008) |
| | - | $97°.91, 6°.66$ | Perryman et al. (1998) |
| $x_c, y_c, z_c$ (pc) | 66.97, 102.81, 56.30 | 18.36 42.20, 13.94 | Present work |
| | 70.63, 103.58, 55.60 | - | Elsanhoury et al. (2018) |
| $S_\odot, l_A, b_A$ | 31.14, -77.47, 26.20 | 46.62, -25.17, 2.89 | Present work |

## 4. Conclusions

In the present paper, morphological analysis for the open clusters Melotte 22 and Melotte 25 is performed using Gaia DR2 and PPMXL catalogues. The analysis was made in the region near infrared J, H, and $K_s$ bands. In what follows we summarize our results:

- The centers of the clusters were calculated for both right ascension and declination, into which for Melotte 22, our right ascension is greater than the value of Kharchenko et al. (2016) by about $8^s.3$ and less than that of Dias et al. (2002) by about $1^m\ 0^s.8$, while the declination is greater than both values of Kharchenko et al. (2016) and Dias et al. (2002) by about $1^m\ 55^s.79$ and $17^m\ 4^s.79$, respectively. On the other hand, for Melotte 25, our right ascension is greater than the values of Kharchenko et al. (2016) and Dias et al. (2002) by about $38^s.24$ and $36^s.24$ respectively, while the declination is greater than the values of Kharchenko et al. (2016) and Dias et al. (2002) by about $7^m\ 55^s.4$ and $7^m\ 56^s.4$ respectively.

- We have calculated the gradient function of the density distribution according to King model (1966), and the results are in a good agreement with that obtained by other authors.

- We have calculated the heliocentric distances, reddening, luminosity, and mass functions by means of isochrones fitting with $Z = 0.019$, log (age) = $8.2 \pm 0.05$ yr for 903 members of Melotte 22 and $Z = 0.024$, log (age) = $8.9 \pm 0.10$ yr for 1712 members of Melotte 25 clusters, and therefore we calculated the masses with aid of mass-luminosity relation MLR, i.e. $662.476 \pm 25.73\ M_\odot$ and $513.819 \pm 22.65\ M_\odot$ for both clusters. Also, we have calculated the projection distances to the Galactic plane.

- According to our estimated ages, crossing and relaxation times, therefore we can estimate the dynamical evolution parameter of the two clusters, which indicate that the clusters are dynamically relaxed.

- The internal motion of members (i.e. kinematics) including of the apex of clusters by means of AD-chart method have been determined.

**Acknowledgments**

The authors would like to thank the referee for very useful and important suggestions that improved this article. This work presents results from the European Space Agency (ESA) space mission, Gaia. Gaia data are being processed by the Gaia Data Processing and Analysis Consortium (DPAC). Funding for the DPAC is provided by national institutions, in particular, the institutions participating in the Gaia MultiLateral Agreement (MLA). The Gaia mission website is https://www.cosmos.esa.int/gaia. The Gaia archive website is https://archives.esac.esa.int/gaia. Virtual observatory tools like http://cds.u-strasbg.fr and TOPCAT were used in the analysis.

We wish to acknowledge the financial support of this work through Northern Border University NBU, deanship of scientific research and higher education grant number SCI-2017-1-8-F-7209.